\newcommand{\pathBibFilesA}{/home/castella/Documents/LatexStyle}
\newcommand{\pathStyleFiles}{.}
\documentclass[onecolumn]{IEEEtran}

\usepackage[utf8]{inputenc}

\usepackage{amsmath}
\usepackage{dsfont}
\usepackage{amssymb}
\usepackage[mathscr]{eucal}
\usepackage{xcolor} 

\newcounter{hypie} \newcounter{hyp}
\newenvironment{hyp}
     {\begin{list}{\textsf{A\arabic{hyp}.}}
                  { \usecounter{hyp}   \setcounter{hyp}{\value{hypie}}
                   \setlength{\listparindent}{0pt}
                   \setlength{\labelwidth}{0pt}
                   \setlength{\leftmargin}{10pt}
                   } }
     {\end{list} \setcounter{hypie}{\value{hyp}}}

\newcounter{rqueie} \newcounter{rque}
\newenvironment{rque}
     {\begin{list}{\textbf{Remark\,\arabic{rque}}:}
                  {\usecounter{rque}\setcounter{rque}{\value{rqueie}}
                }
         }
     {\end{list} \setcounter{rqueie}{\value{rque}}}

\newcommand{\bH}{\mathbf{H}}

\newcommand{\bK}{\mathbf{K}}

\newcommand{\bM}{\mathbf{M}}

\newcommand{\bS}{\mathbf{S}}

\newcommand{\bd}{\mathbf{d}}

\newcommand{\bh}{\mathbf{h}}

\newcommand{\bn}{\mathbf{n}}

\newcommand{\bp}{\mathbf{p}}

\newcommand{\bu}{\mathbf{u}}

\newcommand{\bx}{\mathbf{x}}
\newcommand{\by}{\mathbf{y}}
\newcommand{\bz}{\mathbf{z}}

\newcommand{\balpha}{\boldsymbol{\alpha}}
\newcommand{\bbeta}{\boldsymbol{\beta}}
\newcommand{\bgamma}{\boldsymbol{\gamma}}

\newcommand{\bphi}{\boldsymbol{\phi}}

\newcommand{\ndlr}[1]{{\color{blue} (\textsc{#1})}}

\newcommand{\mJ}{\mathcal{J}}

\newcommand{\mL}{\mathcal{L}}
\newcommand{\mM}{\mathcal{M}}

\newcommand{\mP}{\mathcal{P}}

\newcommand{\NN}{\mathbb{N}}

\newcommand{\RR}{\mathbb{R}}

\newcommand{\defeq}{:=}

\newcommand{\st}{\mathrm{s.t.} \;}   

\newcommand{\tr}{{^\top}} 
 
    \DeclareFontFamily{U}{wncy}{}
    \DeclareFontShape{U}{wncy}{m}{n}{<->wncyr10}{}
    \DeclareSymbolFont{mcy}{U}{wncy}{m}{n}
    \DeclareMathSymbol{\Sh}{\mathord}{mcy}{"58}

\def\qed{\ifmmode\hbox{\hfill\sqb}\else{\ifhmode\unskip\fi%
\nobreak\hfil
\penalty50\hskip1em\null\nobreak\hfil$\blacksquare$
\parfillskip=0pt\finalhyphendemerits=0\endgraf}\fi}

\newcommand{\modif}[1]{#1}
\renewcommand{\ndlr}[1]{}
\usepackage{multirow}
\usepackage{subfig}
\ifCLASSINFOpdf
   \usepackage[pdftex]{graphicx}
   \graphicspath{{../fig_eps/},{./}}  
  \DeclareGraphicsExtensions{.pdf,.jpeg,.png}
\else
  \usepackage[dvips]{graphicx}
  \graphicspath{{../fig_eps/},{./}}
  \DeclareGraphicsExtensions{.eps}
\fi

\makeatletter
\def\ps@headings{%
\def\@oddhead{\mbox{}\scriptsize\rightmark \hfil \thepage}%
\def\@evenhead{\scriptsize\thepage \hfil \leftmark\mbox{}}%
\def\@oddfoot{\mbox{} \hfil}
\def\@evenfoot{\scriptsize DRAFT\hfil \@date}}
\makeatother
\newcommand{\bobs}{\mathbf{d}} 
\newcommand{\obs}{d} 
\newcommand{\mommat}[2]{\bM_{#2}(#1)}
\newcommand{\locmat}[3]{\bM_{#2}^{#3}(#1)}
\newcommand{\RieszF}[2]{\mL_{#2}(#1)}

\begin{document}

\title{Rational Optimization for Nonlinear Reconstruction with
  Approximate $\ell_0$ Penalization}

\author{
        Marc Castella, {\em Member, IEEE}, 
        Jean-Christophe~Pesquet,  {\em Fellow, IEEE},
        and Arthur Marmin

        \thanks{M. Castella (corresponding author) is with SAMOVAR,
          T{\'e}l{\'e}com SudParis, CNRS, Universit{\'e} Paris-Saclay,
          9 rue Charles Fourier, 91011 Evry Cedex, France. E-mail:
          \texttt{marc.castella@telecom-sudparis.eu}.

          J.-C. Pesquet and A. Marmin are with the Center for Visual
          Computing, CentraleSup\'elec, INRIA and Universit{\'e}
          Paris-Saclay, 91192 Gif sur Yvette, France.  E-mail:
          \texttt{jean-christophe@pesquet.eu},
          \texttt{arthur.marmin@centralesupelec.fr}.  } }


\maketitle
\thispagestyle{headings}

\begin{abstract}
  Recovering nonlinearly degraded signal in the presence of noise is a
  challenging problem.
  In this work, this problem is tackled by minimizing the sum of a non
  convex least-squares fit
  criterion 
  and a penalty term. We assume that the nonlinearity of the model can
  be accounted for by a rational function.  In addition, we suppose
  that the signal to be sought is sparse and a rational approximation
  of the $\ell_0$ pseudo-norm thus constitutes a suitable
  penalization.  The resulting composite cost function belongs to the
  broad class of semi-algebraic functions.
  To find a globally optimal solution to such an optimization problem,
  it can be transformed into a generalized moment problem, for which a
  hierarchy of semidefinite programming relaxations can be
  built. Global optimality comes at the expense of an increased
  dimension and, to overcome computational limitations concerning the
  number of involved variables, the structure of the problem has to be
  carefully addressed. A situation of practical interest is when the
  nonlinear model consists of a convolutive transform followed by a
  componentwise nonlinear rational saturation. We then propose to use
  a sparse relaxation able to deal with up to several hundreds of
  optimized variables.  In contrast with the naive approach consisting
  of linearizing the model, our experiments show that the proposed
  approach offers good performance.
\end{abstract}


\ifCLASSOPTIONpeerreview
\begin{center} \bfseries EDICS Category: OPT-NCVX, OPT-SOPT,
  OTH-NSP \end{center}
\fi
\IEEEpeerreviewmaketitle

\section{Introduction}
\label{sec:introduction}



\IEEEPARstart{O}{ver} the last decade, there has been much progress
made in the area of sparse signal recovery. The results and techniques
have spread over a wide range of signal processing applications such
as denoising, source separation, image restoration, or image
reconstruction. Attention has been however mostly focused on linear
observation models, for which many efforts have been dedicated to
solving the associated inverse problems. In the basic setup, a vector
of observations $\bobs$ is available, which is obtained from a
ground-truth signal $\overline{\bx}$ by a linear transformation
$\bH$. It is known that an exact reconstruction of $\overline{\bx}$ is
possible even when the size of the latter is greater than the number
of observations, a fact popularized by the celebrated compressed
sensing theory \cite{Candes_E_2008_ieee-spm_Introduction_cs} which
exploits the structure (i.e. sparsity) of $\overline{\bx}$.

Unfortunately, the linear assumption on the observation model is often
quite inaccurate. For a long time and in many signal processing
applications, attempts have been made in order to deal with more
general nonlinear models. For example, one can mention the pioneering
works undertaken with Volterra models \cite{shetzen}, which may be
useful in some application areas
\cite{dobigeon-spmag2014}. \modif{More recently, the work in
  \cite{yang15_spars_nonlin_regres} has explicitly taken into account
  a nonlinearity, but the reconstruction results hold under
  restrictive assumptions.} Similarly, for many real acquisition
devices, the actual degradation model is not linear as some nonlinear
saturation effects often arise. This situation is closely related to
1-bit compressed sensing \cite{BoufounosBaraniuk-CISS08} and
classification problems. Such nonlinearly distorted convolution models
may also be encountered in blind source separation
\cite{deville-lva15} and neural networks \cite{MacKay-NN}. A
simplified model resulting from a linearization procedure can then be
adopted in order to make the associated mathematical problem
tractable. For example, standard tools in signal processing such as
the Wiener filter are effective mostly in a linear framework. More
specifically, well-known sparse recovery techniques such as LASSO have
been used in a nonlinear context by overlooking the nonlinearity. Some
results have been obtained in this context
\cite{Genzel-ieeeIT17,Genzel-spars17,PlanVershynin-ieeeIT17}, but
methods explicitly taking into account the nonlinearity are likely to
provide better results and are crucially lacking. This paper aims at
providing such a method in this still unexplored area.

A popular approach in many reconstruction problems consists in
minimizing the sum of a data fidelity term and a regularization term
incorporating prior information such as sparsity. In this case, convex
potentials related to the $\ell_1$ norm are often employed as
surrogates to the natural sparsity measure, which is the $\ell_0$
pseudo-norm (count of the number of nonzero components in the
signal). Although some theoretical works have promoted the use of the
$\ell_1$ norm \cite{Candes_E_2008_ieee-spm_Introduction_cs}, its
optimality can only been established under some restrictive
assumptions. In turn, cost functions involving the $\ell_0$
pseudo-norm lead to NP-hard problems for which reaching a global
minimum cannot be guaranteed in general
\cite{Nikolova_M_2013_siam-is_Description_mlsrugm,Blumensath_T_2008_j-four-anal-appl_Iterative_tsa,Patrascu_A_2015_ieee-tac_Random_cdmrco}.
Smooth approximations of the $\ell_0$ pseudo-norm may appear to
provide good alternative solutions
\cite{GemanMcClure85,Chouzenoux_2013_siam-is_Majorize_msair,Florescu_2014_sp_Majorize_mmgmcvip,soubies-siam2015}.
Among the class of possible smoothed $\ell_0$ functions, the
Geman-McClure $\ell_2-\ell_0$ potential was observed to give good
results in a number of applications
\cite{GemanMcClure85,Chouzenoux_2013_siam-is_Majorize_msair,Florescu_2014_sp_Majorize_mmgmcvip}.
Yet, in the recent works
\cite{castella-camsap15,castella-eusipco17,castella-spars17,soubies-siam2015}, 
promising results have been obtained with a non differentiable
function.


Concerning the minimization of the penalized criterion, many efforts
have been undertaken to derive efficient algorithms able to deal with
a large number of variables, while ensuring convergence to a global
minimizer
\cite{Combettes_P_2010_inbook_proximal_smsp,Komodakis_pd,_Boyd_admm}. Many
of the available techniques assume that the observation model is
linear and that the noise has a log-concave likelihood. Then, both the
penalty and the data fit terms are convex and many optimization
techniques may be used. In a more difficult scenario,
a quadratic tangent function can be derived, which makes efficient
majorization-minimization (MM) strategies usable for optimizing
certain penalized criteria 
(see \cite{Chouzenoux_2011_ieee-ip_Majorize_msssoair} for more
details). However, for most of the existing optimization algorithms
(e.g. those based on Majorize-Minimize strategies), only convergence
to a local minimum can be expected and algorithms can get trapped by
undesirable local minima due to the nonconvexity of the criterion. In
our context, none of the two terms of the criterion is convex because
of the nonlinear observation model and because of the chosen
approximation of the $\ell_0$ pseudo-norm. Developing methods with
global convergence properties is therefore a crucial issue, which we
address in this paper.


An approach recently proposed in the optimization community
\cite{lasserre01,laurent09,lasserre-book10,jibetean06} provides
theoretical guarantees of global optimality 
when only polynomial or rational functions are involved. The
minimization problem is recast as a problem of moments, for which a
hierarchy of semidefinite positive programming (SDP) relaxations
asymptotically provides an exact solution. This approach is often
referred to as the Lasserre hierarchy \cite{lasserre01} and its major
advantage is a guaranteed convergence to the global minimum of
the original problem, which can be accessed by solving successive SDP
problems. Alternatively, the problem of global polynomial or rational
minimization can be tackled from the standpoint of sum of squares
(SOS) hierarchy
\cite{laurent09,jibetean06,parrilo03_semid_progr_relax_semial_probl}:
both approaches are linked by duality. One advantage of the moment
approach is the possibility, under some conditions, to extract the
optimal point.

We investigate here the potential offered by these rational
optimization methods for sparse signal recovery from nonlinearly
degraded observations. In the present state of research, the
Lasserre/SOS hierarchies are restricted to small to medium size
problems. In signal processing, one of the main difficulties we face
is the large number of variables which have to be optimized. A
stochastic block-coordinate method has been proposed as a first
solution in one of our previous works \cite{castella-camsap15}:
despite interesting experimental results, global optimality is lost in
this case.

In this work, we propose a novel approach for restoring sparse signals
degraded by a nonlinear model. More precisely, our contributions in
this paper are threefold.
\begin{itemize}
\item First, the proposed approach is able to deal with degradation
  models consisting of a convolution followed by a pointwise
  transform. The latter appears as a rational fraction of the absolute
  value of its input
  argument. 
  The formulation of the problem as a nonconvex optimization also
  allows the use of a Geman-McClure like regularization term.

\item Although SDP relaxations of optimization problems are popular in
  signal processing \cite{Luo-spmag2010}, they usually lead to suboptimal
  solutions.  Our second contribution is to make use of asymptotically
  exact SDP relaxations able to minimize polynomial or rational
  nonconvex functions of several variables.

\item The last contribution of this work is to devise a sparse
  relaxation in the spirit of \cite{bugarin15-MinRatFunc} to cope with
  the resulting rational optimization. Exploiting the specific
  structure of the problem to obtain sparse SDP relaxations plays a
  prominent role in making the Lasserre/SOS hierarchy applicable to
  several hundred of variables as it is common in inverse problems.

\end{itemize}

The remainder of the paper is organized as follows. The considered
model is described in Section \ref{sec:model-criterion}. Section
\ref{sec:polyn-rati-optim} describes the general methodology and
Section \ref{sec:expl-probl-struct} emphasizes the specificities of
our context. Simulations results are provided in Section
\ref{sec:simuls}. Finally, Section \ref{sec:conclu} concludes the
paper.

\textbf{Notation}: The set of polynomials in the indeterminates given
by vector $\bx=(x_1,\dots,x_T) \in \RR^T$ is denoted by
$\RR[\bx]$. For any multi-index
$\balpha= (\alpha_1,\dots,\alpha_T)\in\NN^T$, we define
$\bx^{\balpha}= x_1^{\alpha_1}\dots x_T^{\alpha_T}$ and
$|\balpha|=\alpha_1+\dots+\alpha_n$. Therefore, any polynomial can be
written as a finite sum over multi-indices as follows:
$(\forall \bx \in \RR^T)$
$p(\bx)=\sum_{\balpha} p_{\balpha} \bx^{\balpha}$.  The degree of $p$
will be denoted by $\deg p$.  Such a polynomial can be identified with
the vector of its coefficients $\bp=(p_{\balpha})_{\balpha}$: this
will be used for convenience. Finally, the \modif{lowest} integer upper bound
of any real-valued number is denoted by $\lceil.\rceil$.

\section{Model and criterion}
\label{sec:model-criterion}

\subsection{Observation model}
\label{sec:observ-model}

We consider the problem of recovering a set of unknown samples given
by the vector
$\overline{\bx}\defeq(\overline{x}_1,\dots,\overline{x}_T)\tr$. In our
context, this original signal cannot be measured and we have access
only to some measurements related to the original signal through a
linear transformation followed by some nonlinear effects. More
precisely, the observation model reads
\begin{gather}
  \label{eq:modeleNonLin+bruit}
  \bobs = \bphi(\bH\overline{\bx}) + \bn,
\end{gather}
where the vector $\bobs= (\obs_1,\dots,\obs_T)\tr$ contains the
observation samples, $\bn=(n_1,\dots,n_T)\tr$ is a perturbation noise
vector, $\bH\in\RR^{T\times T}$ is a given matrix, and
$\bphi:\RR^T\to\RR^T$ is a nonlinear function. It is assumed that
$\bphi$ applies componentwise, that is, for every
$\bu\defeq(u_1,\dots,u_T)\tr$, the $t$-th component of $\bphi(\bu)$ is
given by $[\bphi(\bu)]_t=\phi_t(u_t)$, where the real-valued function
$\phi_t$ models a saturation effect as in the top plot of
Figure~\ref{fig:FoncRat}. In this paper, the functions
$(\phi_t)_{1\leq t\leq T}$ are assumed to be known and to be rational,
possibly involving absolute values. Actually,
\modif{$(\phi_t)_{1\leq t\leq T}$ being saturation is only a practical
  example and} other functions could be considered in theory as long
as \modif{ the function $\bphi$ in (\ref{eq:modeleNonLin+bruit}) is
  semi-algebraic} 
(see the comments in Subsection \ref{ssec:semi-alg}).

The linear part in Model (\ref{eq:modeleNonLin+bruit}) can 
typically describe a convolution.
\modif{When the matrix $\bH$ is Toeplitz band as in
  Section~\ref{sec:toepl-struct-splitt}, with values defined by the
  finite impulse response $(h_t)_t$ of a filter, and with vanishing
  boundary conditions, the samples in (\ref{eq:modeleNonLin+bruit})
  indeed stem from a signal given by
\begin{gather}
  \label{eq:modeleFiltre+Nonlin+bruit}
  (\forall t \in \{1,\ldots,T\}) \quad \obs_t = \phi_t(h_t\star
  \overline{x}_t) + n_t.
\end{gather}
In the equation above, $\star$ denotes the sequence convolution and
$\big(n_t)_{1\le t \le T}$ is a realization of an additive random
noise.}  An important contribution in this paper is that this
structure will be exploited in order to reduce the computational cost
of the subsequently proposed global optimization method (see Section
\ref{sec:expl-probl-struct}). For now, no assumption is made on the
matrix $\bH$.

\subsection{Sparse signal and penalized criterion}
\label{sec:sparse-sign-penal}

The signal $(\overline{x}_t)_{1\leq t\leq T}$ modelled by vector
$\overline{\bx}$ is assumed to be sparse. By saying this, we simply
assume that $\overline{x}_t\neq 0$ only for a few indices~$t$.

\begin{figure}[!t]
  \centering
  \includegraphics[width=8.8cm]{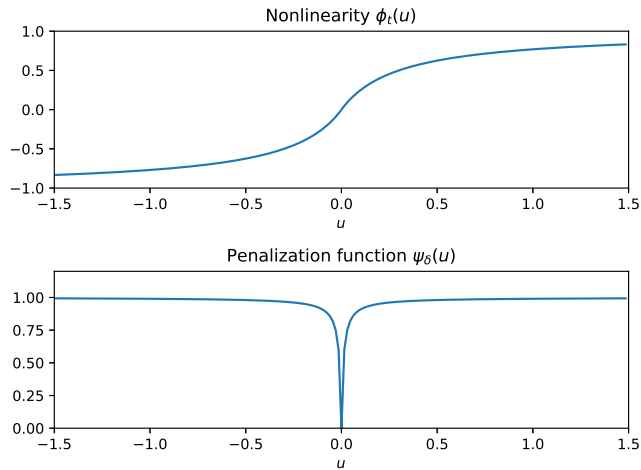}
  \caption{Plot of the nonlinear saturation function $\phi_t$ in
    \eqref{eq:phiNL} with $\chi=0.3$ and of the sparsity
    promoting function in \eqref{eq:psiGMC_abs} with
    $\delta=0.01$.}
  \label{fig:FoncRat}
\end{figure}

Following a classical approach for estimating $\overline{\bx}$, we
minimize a penalized criterion having the following form:
\begin{gather}
  \label{eq:crit_Penal}
  (\forall \bx \in \RR^T)\quad
  \mJ(\bx) = \|\bobs - \bphi(\bH\bx)\|^2 + \lambda\mP(\bx)  \,,
\end{gather}
where $\mP$ is a penalization function whose small values promote
sparse vectors, in accordance with our assumptions concerning the true
$\overline{\bx}$. The positive regularization parameter $\lambda$
controls the relative importance given to the squared norm fit term
and to the penalization. In this paper, we have chosen
$\mP(\bx) = \sum_{t=1}^T \psi_\delta(x_t)$ where the sparsity
promoting function $\psi_\delta$ has been drawn on the bottom plot of
Figure~\ref{fig:FoncRat} and is given by
\begin{equation}
  \label{eq:psiGMC_abs}
  (\forall \xi \in \RR) \qquad \psi_\delta(\xi) = \frac{|\xi|}{\delta+|\xi|} \;.
\end{equation}
This choice is similar in spirit to the Geman-McClure potential
\cite{GemanMcClure85} and, since for every $\xi \in \RR$,
$\lim_{\delta \to 0}\psi_\delta(\xi) =
0$ 
when $\xi = 0$ and $1$ otherwise, the solution to the
$\ell_0$ penalized problem is recovered asymptotically as $\delta
\to 0$ under some technical assumptions (see \cite[Proposition 2]{Chouzenoux_2013_siam-is_Majorize_msair}). 
Note also that this penalty has recently shown to be effective in
image restoration problems (see
\cite{wang18_noncon_optim_point_sourc_local} and references therin).
Finally, the criterion to be minimized in our approach reads:
\begin{gather}
  \label{eq:crit_PenalGeman}
  (\forall \bx \in \RR^T)\quad \mJ(\bx) = \|\bobs - \bphi(\bH\bx)\|^2
  + \lambda\sum_{t=1}^T \psi_\delta(x_t) \,.
\end{gather}
\modif{The minimization is performed over a feasible set $\bK$,
  which is assumed to be compact. This assumption is required later in
  (\ref{eq:momsequence}) and (\ref{eq:def_K}) and it makes no
  restriction since the signal values $\overline{\bx}$
  are bounded in practice.} The optimal cost function value is denoted
by
\begin{gather*}
  \mJ^\star = \inf_{\bx\in\bK} \mJ(\bx) \; ,
\end{gather*}
and the estimated signal generated by our approach is then
$\widehat{\bx} = \arg \min_{\bx\in\bK} \mJ(\bx)$.

Importantly, our model involves rational functions $\bphi$
and $\psi_\delta$.
As a consequence, the criterion $\mJ$
is a rational function of its arguments (possibly with absolute
values). We detail in next section how recent rational and polynomial
optimization techniques apply in this context.

\section{Polynomial and rational optimization}
\label{sec:polyn-rati-optim}

In this section, we explain the basic principles of the methods in
\cite{lasserre01,lasserre-book10} for polynomial and rational
optimization.

\subsection{Global optimization and equivalent problem over measure}

Consider the problem of determining the \emph{global} minimum of a
given lower-semicontinuous function $f$ over a given compact set
$\bK \subset\RR^T$:
\begin{gather}\label{e:fstar}
\mbox{Find}\quad
  f^\star = \inf_{\bx \in \bK} f(\bx).
\end{gather}
We can introduce an optimization problem equivalent to
\eqref{e:fstar}, where the new optimization variable is a measure
belonging to an infinite dimensional space. Following the terminology
from \cite{lasserre-book10}, such problem will be called a
\emph{generalized problem of moments} (GPM). Denoting by $\mP(\bK)$
the set of probability measures suppported on $\bK$, this problem
reads as follows:
\begin{gather}\label{e:fstargpm}
  \mbox{Find}\quad (f^\star)_{\mathrm{gpm}}= \inf_{\mu\in\mP(\bK)}
  \int_{\RR^T} f(\bx)\;d\mu(\bx) \,
  . 
\end{gather}
To see the equivalence between \eqref{e:fstar} and \eqref{e:fstargpm},
note first that $(\forall \bx \in \bK)$ $f(\bx)\geq f^\star$ implies
that $(f^\star)_{\mathrm{gpm}}\geq f^\star$. For the reverse
inequality, it can be noticed that the minimum of $f$ is reached at a
point $\bx^\star\in \bK$ because $\bK$ is compact and the Dirac
measure $\delta_{\bx^\star}$ provides a solution such that
$(f^\star)_{\mathrm{gpm}}=f^\star$.

Let us now write more specifically the GPM for rational functions by
assuming that the function $f$ reads:
\begin{equation}\label{e:frat}
  (\forall \bx \in \RR^T)\;\; f(\bx)=\frac{p(\bx)}{q(\bx)}
  \quad \mbox{where}\quad  (\forall \bx \in \bK)\;\; q(\bx)>0 \,,
\end{equation}
\modif{where $p$ and $q$ are polynomials.}
Let us introduce the measure
$d\nu(\bx)=\frac{1}{q(\bx)}d\mu(\bx)$. With this change of variables,
$\nu$ is no longer a probability measure, but, since the total mass of
the probability measure $\mu$ is one, it satisfies
(\ref{eq:fstar_gpm_2}) below. Therefore, by defining $\mM(\bK)$ as the
set of finite nonnegative Borel measures supported on $\bK$,
Problem~\eqref{e:fstargpm} can be equivalently re-expressed as:
\begin{align}
\mbox{Find}\quad
  \label{eq:fstar_gpm}
  f^\star= & \inf_{\nu\in\mM(\bK)} \int_{\RR^T} p(\bx)\;d\nu(\bx) \\ 
  & \st \int_{\RR^T} q(\bx) d\nu(\bx) = 1 \, . \label{eq:fstar_gpm_2}
\end{align}
Importantly, Problem \eqref{eq:fstar_gpm}-\eqref{eq:fstar_gpm_2}
corresponds to a simple objective function and an explicit constraint,
both terms being linear with respect to $\nu$. However, the implicit
contraint that $\nu$ is a measure in $\mM(\bK)$ is more complicated to
cope with. Fortunately, the latter can be handled via a hierarchy of
tractable constraints when $p$ and $q$ are polynomials, as shown next.

\subsection{Hierarchy of SDP relaxations}
\label{sec:hier-sdp-relax}

The infinite dimensional optimization problem
$\eqref{eq:fstar_gpm}-\eqref{eq:fstar_gpm_2}$ can be approximated by a
hierarchy of SDP problems with increasing sizes when the involved
function is given by \eqref{e:frat} with $(p, q) \in(\RR[\bx])^2$. The
main ingredients of this approach are presented now.

\subsubsection{Moment sequence}
In \eqref{eq:fstar_gpm}-\eqref{eq:fstar_gpm_2}, the optimization
variable is the measure $\nu$. The first step is to replace this
variable by a more tractable one, i.e. a finite dimensional
vector. Since the measure $\nu$ has a compact support, it can be
represented by a moment sequence
$\by = (y_{\balpha})_{\balpha\in\NN^T}$ defined as
\begin{gather}
  \label{eq:momsequence}
 (\forall \balpha \in \NN^T)\quad y_{\balpha}= \int_{\bK} \bx^{\balpha} \;d\nu(\bx).
\end{gather}
In so doing, the measure $\nu$ in Problem
\eqref{eq:fstar_gpm}-\eqref{eq:fstar_gpm_2} is represented by the
moment sequence $\by$, which is an infinite dimensional vector.
A hierarchy of finite dimensional optimization problems will be
obtained by considering truncated versions of $\by$ with increasing
sizes.

\subsubsection{Linear objective and constraints}

Consider a polynomial of total degree $k$ represented by its vector of
coefficients $\bp=(p_{\balpha})_{|\balpha|\leq k}$:
\begin{gather}
  \label{eq:poly_p}
  (\forall \bx \in \RR^T)\quad  p(\bx)=\sum_{|\balpha|\leq k} p_{\balpha}\bx^{\balpha}.
\end{gather}

By linearity and by the definition of the moments
$(y_{\balpha})_{\balpha}$, any integral such as the ones arising in
(\ref{eq:fstar_gpm}) and (\ref{eq:fstar_gpm_2}) can be rewritten as
\begin{gather}
  \label{eq:Lp(y)}
  \int_{\RR^T} p(\bx) \;d\nu(\bx) 
  =  \sum_{|\balpha|\leq k} p_{\balpha} y_{\balpha} = \RieszF{\by}{p}.
\end{gather}
The function $\RieszF{\cdot}{p}$ as defined above is
linear. Therefore, the objective and the explicit constraint in
(\ref{eq:fstar_gpm}) and (\ref{eq:fstar_gpm_2}) are linear funtions of
the moment sequence $\by$ and the difficulty of the original problem
has therefore been transfered to the implicit constraint that the
sequence $\by$ should satisfy (\ref{eq:momsequence}) for a given
measure $\nu\in \mM(\bK)$.



\subsubsection{Support/measure constraint}

Since an arbitrary sequence $\by$ does not necessarily represent a
measure $\nu$ on $\bK$, some constraints needs to be taken into
account on $\by$. To achieve this goal, we first need a more precise
description of $\bK$. In our context, the set $\bK$ is defined by
polynomial inequalities of the following form:
\begin{gather}
  \label{eq:def_K}
  \bK = \{\bx\in\RR^T\mid (\forall i\in\{1,\dots,I\})\,g_i(\bx)\geq 0\},
\end{gather}
where, for every $i\in \{1,\ldots,I\}$, $g_i \in \RR[\bx]$. The
constraints will now be specified on a truncated version of the
sequence $\by$. For a given $k\in\NN$ and for multi-indices
$\balpha,\bbeta$ such that $|\balpha| \leq k$ and $|\bbeta|\leq k$,
the elements of the $k$-th order \emph{moment matrix}
$\mommat{\by}{k}$ of $\by$ are given by
\begin{gather}
  \label{eq:momentMatrix}
  \left[ \mommat{\by}{k} \right ]_{\balpha,\bbeta} = y_{\balpha+\bbeta}.
\end{gather}
Note that $\mommat{\by}{k}$ 
involves moments up to the order $2k$. The main property of
$\mommat{\by}{k}$ is that for a polynomial of degree no more than $k$
expressed by (\ref{eq:poly_p}), we have:
\begin{gather}
  \label{eq:int_f2_equals_ftMf}
  \int_\bK p(\bx)^2\;d\nu(\bx) = \bp\tr \mommat{\by}{k} \bp. 
\end{gather}
Similarly, for a given polynomial $g\in\RR[\bx]$, the elements of the
\emph{localizing matrix} $\locmat{\by}{k}{g}$ associated to $g$ and
$\by$ are
\begin{gather}
  \label{eq:localizingMatrix}
  \left[ \locmat{\by}{k}{g} \right ]_{\balpha,\bbeta} = \sum_{\bgamma}
  g_{\bgamma}y_{\bgamma+\balpha+\bbeta},
\end{gather}
and we have
\begin{gather}
  \label{eq:int_gf2_equals_ftMf}
    \int_\bK g(\bx)p(\bx)^2\;d\nu(\bx) = \bp\tr \locmat{\by}{k}{g} \bp.
\end{gather}
The positivity of the right hand side of (\ref{eq:int_f2_equals_ftMf})
for any vector of coefficients $\bp$ shows that the positive
semi-definiteness of matrix $\mommat{\by}{k}$ is a necessary condition
for the sequence $\by$ to be a valid moment sequence. Similarly,
according to (\ref{eq:int_gf2_equals_ftMf}) and because
$(\forall \bx \in \bK)$ $g_i(\bx)\geq 0$, we deduce that
$\locmat{\by}{k}{g_i}\succeq 0$ for every $i\in \{1,\dots,I\}$, if
$\by$ is the moment sequence of a measure in $ \mM(\bK)$. Due to the
linear dependence of $\mommat{\by}{k}$ and $\locmat{\by}{k}{g}$ on
$\by$, these constraints are linear matrix inequalities.

\subsubsection{Relaxation}
Based on the above developments, we are now able to introduce a
relaxation of Problem
\eqref{eq:fstar_gpm}-\eqref{eq:fstar_gpm_2}. Define, for every
$i\in\{1,\dots,I\}$, $r_i= \lceil (\deg g_i)/2 \rceil$ and, for any
order $k\geq \max \{\max_{i=1}^I r_i, \deg p, \deg q \}$, consider the
optimization problem:
\begin{gather}
\label{eq:Pstar_k}
    \begin{aligned}
    \mbox{Find}\quad
      f^\star_k = \inf_\by \; &  \RieszF{\by}{p} \\
      \st \;&  \RieszF{\by}{q} = 1, \\
      & \mommat{\by}{k} \succeq 0,  \\
      & \locmat{\by}{k-r_i}{g_i} \succeq 0 \quad (\forall i\in\{1,\dots,I\}).
    \end{aligned}
\end{gather}
The objective function and the equality constraint are directly
derived from Problem \eqref{eq:fstar_gpm}-\eqref{eq:fstar_gpm_2} where
\modif{the} integrals have been represented as in
(\ref{eq:Lp(y)}). The last two constraints are necessary constraints
for $\by$ to be a measure supported by $\bK$. Therefore, we naturally
have $f^\star_k\leq f^\star$ and $f^\star_k$ is an increasing sequence
with lower bounds of $f^\star$. Note that since the order in the last
constraints have been limited to $k-r_i$, for every
$i\in \{1,\ldots,I\}$, it follows from \eqref{eq:momentMatrix} and
\eqref{eq:localizingMatrix} that the moments involved in Problem
\eqref{eq:Pstar_k} are $(y_{\balpha})_{|\balpha|\leq 2k}$.

A crucial observation is that \eqref{eq:Pstar_k} is a convex SDP
optimization problem for which efficient techniques exist and provide
guaranteed global optimal solution \cite{Tutuncu2003_sdpt3,Anjos2012}.

\subsubsection{Theoretical results and solution extraction}
\label{sec:theor-results-solut}

\paragraph{Convergence results}
We now detail some existing theoretical results about the
approach. For their validity, the following technical assumption is
required:
\begin{hyp}
\item \label{hyp:algCompact}
  There exist polynomials $\sigma_0,\dots,\sigma_I$, which are all sum of
  squares, such that the set $\{\bx\in\RR^T\,|\, \sigma_0(\bx) +
  \sum_{i=1}^I\sigma_i(\bx)g_i(\bx) \geq 0 \}$ is compact.
\end{hyp}
Under Assumption \ref{hyp:algCompact}, we have
\cite{lasserre01,lasserre-book10}
\begin{gather*}
  f^\star_k \uparrow f^\star \text{ as } k\to+\infty \;.
\end{gather*}
This is a strong result ensuring convergence to the \emph{global}
optimum of Problem \eqref{e:fstar} when considering increasing order
SDP relaxations.

Note that, in addition to $\bK$ being compact, Condition
\ref{hyp:algCompact} requires that that the polynomials
$(g_i)_{1\le i \le I}$ describing $\bK$ in (\ref{eq:def_K}) yield an
algebraic certificate of compactness. More details can be found in
\cite{laurent09,lasserre-book10,jibetean06,bugarin15-MinRatFunc}. For
simplicity, we will only consider the practical situation where
$\bK=[\underline{B},\overline{B}]^T$. This is easily satisfied when
lower and upper bounds $\underline{B},\overline{B}$ on the variables
$(\overline{x_t})_{1\leq t\leq T}$ are available. By setting $I=T$ and
\begin{gather*}
  (\forall \bx\in \RR^T)\qquad g_t(\bx)=
  (x_t-\underline{B})(\overline{B}-x_t),
\end{gather*}
$\bK$ can be expressed under the form \eqref{eq:def_K}.  The set $\bK$
is obviously compact.  In addition,
\begin{align}
  \sum_{t=1}^T g_t(x_t) \ge 0
  \;\;\Leftrightarrow\;\; 
  & \|\bx\|^2 - (\overline{B}+\underline{B}) \sum_{t=1}^T x_t + T \underline{B}\overline{B} \le 0 \nonumber \\
  \Leftrightarrow\;\;  
  & \|\bx-\bu\| \le T \frac{\overline{B}-\underline{B}}{2},
\end{align}
where
$\bu = \frac{\overline{B}+\underline{B}}{2}(1,\ldots,1)^\top \in
\mathbb{R}^T$.
Therefore, Assumption~\ref{hyp:algCompact} holds with
$\sigma_0(\bx) = 0$ and, for every $t\in \{1,\ldots,T\}$
$\sigma_t(\bx) = 1$.

\paragraph{Extraction of the optimal solution} The above convergence
results are asymptotic results for increasing orders of the hierarchy
of SDP relaxations. Fortunately, it has been experimentally observed
that low relaxation orders often provide satisfactory results (see
e.g. \cite{lasserre01,bugarin15-MinRatFunc}). In addition, it has been
proven \cite{lasserre-book10,laurent09} that under certain rank
conditions, the solution given by the SDP relaxation is guaranteed to
be the global minimizer of the original problem. In this case,
globally optimal points can be extracted by the procedure in
\cite{henrion05-bookchap}. Details on the rank conditions and the
extraction procedure are beyond the scope of this paper.

Unfortunately, there are two main difficulties in applying this
methodology to practical situations: first, it is known that detecting
the rank of a matrix can be numerically sensitive. In addition,
because of the complexity of the original problem, the possible
relaxation order that we can choose may be too small. For both
reasons, we have observed that the mentioned rank conditions are
generally not satisfied numerically. Alternatively, considering that
the global minimium is likely to be unique, one can extract from the
optimal solution $\by^\star$ to Problem \eqref{eq:Pstar_k} the moments
corresponding to the respective monomials $x_1,\dots,x_T$. This
extraction is straightforward and we have used the vector of these
moments as an estimate denoted by $\hat{\bx}_k^\star$ of the global
minimizer for the original problem.

\subsubsection{Extension to semi-algebraic functions/constraints}
\label{ssec:semi-alg}

From a theoretical viewpoint, the above methodology can be extended to
more complicated functions and constraints than polynomials or
fractions. We briefly explain how the absolute value, which appears in
the nonlinearity and/or in the penalty function, can be handled.
Proceeding similarly, it is actually possible to handle any
semi-algebraic function or constraint.

First, note that polynomial equality constraints such as $g(\bx)=0$
are possible in the definition of the feasible set $\bK$. This is
easily done by introducing the two inequalities $g(\bx)\geq 0$ and
$-g(\bx)\geq 0$ in the equations defining $\bK$ in (\ref{eq:def_K}).

Then, absolute values can be considered as follows: for each term
$|v(\bx)|$ appearing, where $v$ is a polynomial, one can introduce an
additional variable $u$ and impose the constraints
$u\geq 0, u^2= v(\bx)^2$. The methodology of the paper can then be
applied with the extended set of variables $(\bx,u)$.

\section{Exploiting the problem structure}
\label{sec:expl-probl-struct}

\subsection{Toeplitz structure and split criterion}
\label{sec:toepl-struct-splitt}

In this section, we assume that the convolutional model in
(\ref{eq:modeleFiltre+Nonlin+bruit}) is considered. Additionally, it
is assumed that the involved filter is FIR with impulse response of
length $L$ given by the vector $(h_1,\dots,h_L)\tr$. Under vanishing
boundary conditions, the observation model in
(\ref{eq:modeleNonLin+bruit}) holds and involves the following
specific Toeplitz band matrix:
\begin{gather*}
  \bH = 
  \begin{bmatrix}
    h_1    & 0      & \hdotsfor{3}             & 0      \\
    \vdots & \ddots & \ddots &        &        & \vdots \\
    h_L    &        &        &  \ddots&        & \vdots  \\
    0      & \ddots &        &        &  \ddots& \vdots \\ 
    \vdots & \ddots & \ddots &        &  \ddots&  0    \\
    0      & \hdots & 0      & h_L    & \hdots & h_1
  \end{bmatrix} \,.
\end{gather*}
Finally, remind that we have assumed that the nonlinearity $\bphi$
applies componentwise and that it is given by a rational function,
possibly involving absolute values. The latters can be discarded by
using the trick described above. Thus, for clarity, and with no loss
of generality, we describe the method when all quantities are
nonnegative and hence no absolute value appears.

We now focus on two specificities of our problem and show how they can
leverage a methodology similar to \cite{bugarin15-MinRatFunc}.
\modif{Note that the methodology remains applicable when a subsampling
  is performed on the observation vector $\by$ (see
  \cite{marmin-eusipco18}).} First, developing the squared norm in
(\ref{eq:crit_PenalGeman}) and substituting all terms, the criterion
$\mJ$ appears as a sum of rational functions. Reducing $\mJ$ to the
same denominator would result in a ratio of high degree polynomials,
making the approach described in Section \ref{sec:polyn-rati-optim}
intractable. A remedy consists in introducing one measure (and hence
one moment sequence) for each elementary fraction in $\mJ$, and
simultaneously imposing constraints which guarantee equality of
identical moments related to different measures.

Going further, the second specificity stems from the Toeplitz band
structure of the matrix $\bH$. In this case indeed, each term of the
sum of rational functions in $\mJ(\bx)$ only involves a small subset
of all variables. This leads to a sparse\footnote{The notion of
  sparsity here concerns the optimization variables and should not be
  confused with the sparsity assumed for the original samples in
  vector~$\bx$.}  SDP relaxation. The rationale is explained below and
more details are given in Section \ref{sec:sparse-sdp-relax}.

Let us introduce, for every $t\in\{1,\dots,T\}$, the set
\begin{gather*}
  I_t = \{ \min\{1,t-L+1\}, \dots, t \}
\end{gather*}
which is the set of column indices where $t$-th row of $\bH$ has
nonzero elements (in particular, $I_1=\{1\}$, $I_2=\{1,2\}$, \dots,
$I_T=\{T-L+1,\dots,T\}$). Developing the squared norm, we rewrite
Criterion (\ref{eq:crit_PenalGeman}) as follows
\begin{gather}
  \label{eq:critereJ_separe1}
  \mJ(\bx) = \sum_{t=1}^T \underbrace{\left( \obs_t -
      \phi_t\Big(\sum_{i=1}^L h_ix_{t-i+1} \Big) \right)^2
  }_{\mbox{$\frac{p_{I_t}(\bx)}{q_{I_t}(\bx)}$}} + \underbrace{\lambda
    \psi_\delta(x_t)}_{\mbox{$\frac{p(x_t)}{q(x_t)}$}},
\end{gather}
where by convention $x_t=0$ for every $t\notin\{1,\dots,T\}$. This
reads equivalently:
\begin{gather}
  \label{eq:critereJ_separe2}
  \mJ(\bx) = \sum_{t=1}^T \left( \frac{p_{I_t}(\bx)}{q_{I_t}(\bx)} +
    \frac{p(x_t)}{q(x_t)} \right).
\end{gather}
In the above equation, $p_{I_t}, q_{I_t}$ are polynomials that depend
on the variables $(x_k)_{k\in I_t}$ only and $p(x_t), q(x_t)$ are
univariate polynomials that depend on $x_t$ only.

Now, one can see that, by introducing for each fraction summing up in
(\ref{eq:critereJ_separe1}) a relaxation similar to the methodology
introduced in Section \ref{sec:polyn-rati-optim}, the original problem
involving a large number $T$ of variables is split in a collection of
smaller problems and relaxations. Proceeding in this way would be
quite natural for a separable criterion where the problem is
decomposed into a sum of subproblems that can be solved
independently. Of course, for a non separable criterion, one cannot
split freely the problem and constraints must be added between the
subproblems to link them. In addition, a technical condition is
required on the subsets of variables of the split form. This is
further explained in the next section.

\subsection{Sparse SDP relaxation}
\label{sec:sparse-sdp-relax}
For every $t\in \{1,\ldots,T\}$, each rational function
$\frac{p_{I_t}(\bx)}{q_{I_t}(\bx)}$ is related to the marginal
$\mu_{I_t}$ on $\RR^{| I_t |}$ of the original probability measure
$\mu$ defined on $\RR^T$. By weighting $\mu_{I_t}$ with the
denominator of this rational fraction, as explained in
Section~\ref{sec:polyn-rati-optim}, we define a measure $\nu_{I_t}$
associated with a sequence of moments $\bz_t$, which satisfies the
following relations: for any
$k\geq \max \{1, \deg p_{I_t}, \deg q_{I_t} \}$,
\begin{equation}\label{e:condzt}
  \mommat{\bz_t}{k} \succeq 0,\;\;  
  \RieszF{\bz_t}{q_{I_t}} = 1,\;\;  
  \locmat{\bz_t}{k-r_t}{g_t} \succeq 0.
 \end{equation}
 In addition, we have to pay attention to the fact that the same
 monomial may appear \modif{in consecutive} terms
 $\frac{p_{I_{t-1}}(\bx)}{q_{I_{t-1}}(\bx)}$ and
 $\frac{p_{I_t}(\bx)}{q_{I_t}(\bx)}$ in Summation
 (\ref{eq:critereJ_separe2}), when $t\in \{2,\ldots,T\}$.  Let
 $\NN^{(I_t\cap I_{t-1})}$ denote the subset of $T$-tuples
 $\balpha=(\alpha_1,\dots,\alpha_T) \in \NN^T$ such that $\alpha_t=0$
 for $t\notin I_t\cap I_{t-1}$. In other words, the $T$-tuples in
 $\NN^{(I_t\cap I_{t-1})}$ correspond to monomials involving variables
 with indices in $I_t\cap I_{t-1}$. The latter monomials are precisely
 the common monomials in $\frac{p_{I_{t-1}}(\bx)}{q_{I_{t-1}}(\bx)}$
 and $\frac{p_{I_t}(\bx)}{q_{I_t}(\bx)}$ .  We have then, for every
 $\balpha\in\NN^{(I_t\cap I_{t-1})}$,
\begin{align}
  \label{e:condztzt-1}
  & \int \bx^{\balpha} d\mu_{I_t}(\bx) = \int \bx^{\balpha}  d\mu_{I_{t-1}}(\bx) 
    \nonumber\\
  \Leftrightarrow\quad    
  & \RieszF{\bz_t}{\bx^{\balpha}q_{I_t}(\bx)} = 
    \RieszF{\bz_{t-1}}{\bx^{\balpha}q_{I_{t-1}}(\bx)}.
\end{align}
Similarly, for every $t\in \{1,\ldots,T\}$, the rational function
$\frac{p(x_t)}{q(x_t)}$ can be associated with a sequence of
monovariate moments $\by_t$, for which the following conditions have
to be met:
\begin{equation}\label{e:condyt}
  \mommat{\by_t}{k} \succeq 0,\;\; 
  \RieszF{\by_t}{q_t} = 1,\;\; 
  \locmat{\by_t}{k-r_t}{g_t} \succeq 0, 
\end{equation}
and, for every $\alpha\in \NN$,
\begin{equation}\label{e:condytzt}
  \RieszF{\by_t}{ x_t^\alpha q(x_t)} = \RieszF{\bz_t}{ x_t^\alpha q_{I_t}(\bx) }. 
\end{equation}
By using these variables $(\by_t, \bz_t)_{1 \le t \le T}$, we are now
in order to provide a sparse SDP relaxation for the minimization of
(\ref{eq:critereJ_separe2}):
\begin{gather*}
  \begin{aligned}
    \mbox{Find}\;
    f^{\star\mathsf{s}}_k = &\inf_{\bz,\by} \;   \sum_{t=1}^T \RieszF{\bz_t}{p_{I_t}} + \RieszF{\by_t}{p} \\
    \st \; & (\forall t\in\{1,\dots,T\}): \\
    & \eqref{e:condzt}, \eqref{e:condyt},\\
    & \eqref{e:condztzt-1} \text{ for } \balpha\in\NN^{(I_t\cap
      I_{t-1})} \text{ with } |\balpha| + \deg q_{I_t} \leq 2k, \\
    & \eqref{e:condytzt}  \text{ for } \alpha + \deg q_{I_t} \leq 2k.
  \end{aligned}
\end{gather*}

\begin{rque}
\item For the aforementioned approach to be mathematically valid, a
  technical assumption is required: the so-called Running Intersection
  Property \cite{bugarin15-MinRatFunc,lasserre-book10}.  For
  convenience, let us introduce a notation for the $2T$ different
  index sets corresponding to each fraction in
  (\ref{eq:critereJ_separe2}):
  \begin{gather*}
    (\forall t \in \{1,\dots,T\})\quad J_t=I_t \text{ and } J_{t+T} =
    \{t\}.
  \end{gather*}
  Note that the sets $(J_t)_{1\le t\le 2T}$ satisfy
  $\bigcup_{t=1}^{2T} J_t = \{1,\dots,T\}$ . The Running Intersection
  Property then reads
  \begin{gather}
    (\forall t \in \{2,\ldots,2T\})\;
    J_t\bigcap\left(\bigcup_{k=1}^{t-1}J_k\right) \subseteq J_j 
    \text{ for some $j\leq t-1$}.
  \end{gather}
  It is easy to check that this condition is satisfied in our case.
\end{rque}

\subsection{Comparison between full and sparse relaxations}
\label{sec:comp-betw-full-spars-relax}

We detail here the reasons why the specific form of the latter
relaxation is crucial from a computational standpoint. Using the
sparse relaxation indeed allows us to handle a much higher number of
variables $T$ than the non sparse one. The different numbers of
involved variables and matrix sizes are listed below, in the case when 
no absolute value appears.
\subsubsection{Relaxation involving one measure only}
\label{sec:relax-involv-one}
For a problem with $T$ variables and a relaxation order $k$, the size
of the vector representing the measure/moment sequence is given by the
number of all mutivariate monomials with degree less than or equal to
$2k$, which is precisely the binomial coefficient
$\binom{T+2k}{2k}$. As a consequence, the number of variables in an
SDP relaxation involving only one measure (such as \eqref{eq:Pstar_k})
scales as $T^{2k}$. In addition, according to the definition of the
moment matrix in (\ref{eq:momentMatrix}), the maximum size of the
square matrices defining positive definite constraints is
$\binom{T+k}{k}$, which scales as $T^k$.
\subsubsection{Sparse relaxation for a Toeplitz matrix} Concerning the
sparse relaxation with order $k$, the number of variables involved is
$\binom{L+2k}{2k}$ for each $\bz_t$ and $\binom{1+2k}{2k}=2k+1$ for
each $\by_t$ with $t\in\{1,\ldots,T\}$. The total number of variables
in the sparse SDP relaxation is therefore
\begin{gather*}
  T\left( \binom{L+2k}{2k} + 2k+1 \right).
\end{gather*}
As a consequence, for a given order $k,$ the number of variables
scales as $T\,L^{2k}$ in the computation of
$f^{\star\mathsf{s}}_k$. The maximum size of the moment matrix with
positive definite constraint is then $\binom{L+k}{k}$, hence it scales
as $L^k$.

In summary, the gain in terms of size of the sparse relaxation is
$T^{2k-1}/L^{2k}$. In addition, the maximum size of the semidefinite
constraints is of the respective order $T^k$ for the non sparse
relaxation and $L^k$ for the sparse one. Considering these two facts,
it follows that the sparse relaxation is highly advantageous for
$L\ll T$, that is for $\bH$ corresponding to a convolutive matrix with
relatively short FIR.

\begin{rque}
\item The relaxation order $k$ must be greater than or equal to the
  maximal degree appearing in the original polynomial or rational
  problem. Consequently, Relaxation~\eqref{eq:Pstar_k} is intractable
  after reducing the terms in \eqref{eq:critereJ_separe2} to the same
  denominator, since this would introduce polynomials with a high
  degree (of order $T$). On the contrary, the sparse relaxation takes
  explicitly into account that the criterion is a sum of fractions
  with low degrees and allows order $k$ to be set to a much smaller
  value.
\end{rque}

\section{Simulations}
\label{sec:simuls}

\subsection{Experimental setup}

\subsubsection{Generated sparse signal and nonlinearity}

In all the performed experiments, several sets of $100$ Monte-Carlo
realizations of generated data are processed.  Samples
$\overline{\bx}$ of a sparse signal are generated, the number of
samples being set to $T=200$, $T=100$, $T=50$, or $T=20$. We impose
that exactly 10\% of the sample values are nonzero, yielding
respectively 20, 10, 5, and 2 nonzero components in $\overline{\bx}$.

Then, this impulsive signal, considered as the ground truth, is
corrupted following the model in (\ref{eq:modeleNonLin+bruit}), where
the noise $\bn$ is drawn according to an i.i.d. zero-mean Gaussian
distribution with standard deviation $\sigma = 0.15$. The components
of the nonlinear function $\bphi$ are chosen all identical and given
by
\begin{gather}
  \label{eq:phiNL}
  (\forall t\in\{1,\dots,T\}) \quad \phi_t(u) = \frac{u}{\chi + |u|},
\end{gather}
where $\chi = 0.3$. Considering the amplitude of the signals, the
above function acts as a nonlinear saturation (see top plot in
Figure~\ref{fig:FoncRat}).
Finally, the matrix $\bH$ is Toeplitz band and corresponds to FIR
filters of length 3.

We test our approach in two scenarios:
\paragraph{Nonnegative case} We first consider only a nonnegative
original signal $\overline{\bx}$ and nonnegative coefficients in the
matrix $\bH$. In the Geman-McClure penalty term given by
(\ref{eq:psiGMC_abs}), absolute value are then of no use and they can
be discarded. The amplitudes of the nonzero components of
$\overline{\bx}$ are drawn according to a uniform distribution on
$[2/3,1]$. The impulse responses of the FIR filters corresponding to
$\bH$ are set to $\bh^{(a)}=[0.1, 0.8, 0.1]$ or
$\bh^{(b)}=[0.2254, 0.3361, 0.4385]$.
An additional set of Monte-Carlo simulations is run where the impulse
responses are drawn randomly (nonnegatively) for each realization. Due
to the positivity assumption, the minimization of $\mJ^\star$ is then
performed on the hypercube $\bK=[0,1]^T$.

\paragraph{Real-valued case} We then consider real valued
$\overline{\bx}$ and $\bH$, still using the penalty term in
(\ref{eq:psiGMC_abs}).  The amplitudes of the nonzero components of
$\overline{\bx}$ are then drawn according to a uniform distribution on
$[-1,-2/3]\cup[2/3,1]$. In addition, the impulse responses of the FIR
filters are given by $\bh^{(a)}$, $\bh^{(b)}$ (like in the first
scenario), and $\bh^{(c)}=[-0.1127, -0.0683, 0.8191 ]$.
Here again, on one set of Monte-Carlo realizations, the impulse
responses are randomly drawn with real-valued coeffcients, for each
realization. Finally, the criterion is minimized on the set
$\bK=[-1,1]^T$.

\subsubsection{Considered optimization methods}

Recall that the optimized criterion is given by
(\ref{eq:crit_PenalGeman}). In both scenarios, we have set empirically
$\lambda=0.15$ for the regularization parameter and $\delta=0.01$ in
the penalty function (\ref{eq:psiGMC_abs}).

To obtain an estimate of $\overline{\bx}$, we have built the sparse
SDP relaxation from Section \ref{sec:sparse-sdp-relax} with orders
$k=2$ and $k=3$ using the software \cite{henrion09}. The SDP has then
been solved using SDPT3 \cite{Tutuncu2003_sdpt3}. Finally, the
corresponding estimate $\bx^{\star \mathsf{s}}_k$ is determined as
described in Section \ref{sec:hier-sdp-relax}.

We are not aware of any other method able to find the global minimum
of (\ref{eq:crit_PenalGeman}). For comparison with a globally
convergent approach, we have used a linearized model for
reconstruction purposes: based on Model (\ref{eq:modeleNonLin+bruit}),
we have linearized around zero the nonlinearity (\ref{eq:phiNL}) and
have used the well-known
$\ell_1$ penalization. The cost function then reads
\begin{gather*}
  (\forall \bx \in \RR^T)\quad \mJ_{\ell_1}(\bx) = \Big\|\bobs -
  \frac{1}{\chi} \bH\bx\Big\|^2 + \lambda_1\sum_{t=1}^T
  |x_t|,\;\lambda_1 >0
\end{gather*}
and it can thus be minimized efficiently by standard convex
optimization techniques
\cite{Combettes_P_2010_inbook_proximal_smsp,chaux07_variat_formul_frame_based_inver_probl}.

Finally, we have also implemented a proximal gradient algorithm
corresponding to the well-known Iterative Hard Thresholding (IHT)
\cite{Blumensath_T_2008_j-four-anal-appl_Iterative_tsa}. Since the
function $\bphi$ is Lipschitz-differentiable, 
the standard IHT algorithm can be extended to the nonlinear
observation model. This leads to the following iterative algorithm:
\begin{gather*}
  (\forall n \in \mathbb{N}) \;\; \bx^{(n+1)} = 
  \bS_{\sqrt{\lambda_0\eta}} \Big(\bx^{(n)}-\eta
  \bH^\top\nabla \bphi(\bH\bx^{(n)}) \big(\bphi(\bH \bx^{(n)}\big)-\bd)\Big)
\end{gather*}
where the Jacobian matrix $\nabla \bphi(\bH\bx^{(n)})$ is diagonal and
$\bS_{\sqrt{\lambda_0\eta}}$ is the hard thresholder with threshold
value $\sqrt{\lambda_0\eta}$, $\lambda_0 > 0$. It can be shown that
any value of the stepsize $\eta$ in $]0,\eta_{\max}]$ is valid, where
$1/\eta_{\max}= \|\bH\|_{\rm S}^2(1+2\max_{1\leq t\leq T\}} |d_t|)/\chi^2$ is a
Lipschitz constant of the above gradient term ($\|\bH\|_{\rm S}$
denotes the spectral norm of $\bH$).
The latter algorithm however only certifies convergence to a local
minimum of the criterion \cite{Attouch2013}. Due to non convexity, the
local minima are likely to differ from the global minimum
\cite{Nikolova_M_2013_siam-is_Description_mlsrugm}.

\subsection{Results}

\subsubsection{Performance of the proposed relaxation \ndlr{focused on nonnegative case mainly... parce que les résultats sont plus présentables et on ne le dit pas trop fort.}}

\begin{figure}[!t]
  \centering
  \includegraphics[width=8.8cm]{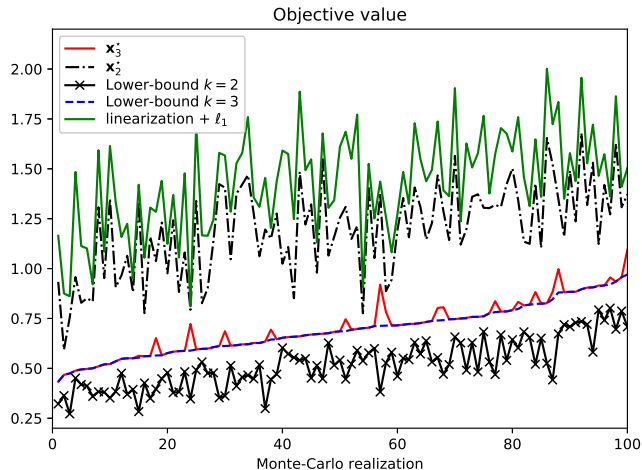}
  \caption{Objective value and lower-bound given by our method
    (randomly driven filters, nonnegative case,
    $T=20$). \ndlr{NLInv\_Prog/Simu2.4\_Pos/}}
  \label{fig:Obj_OrdT20}
\end{figure}

\begin{figure}[!t]
  \centering
  \includegraphics[width=8.8cm]{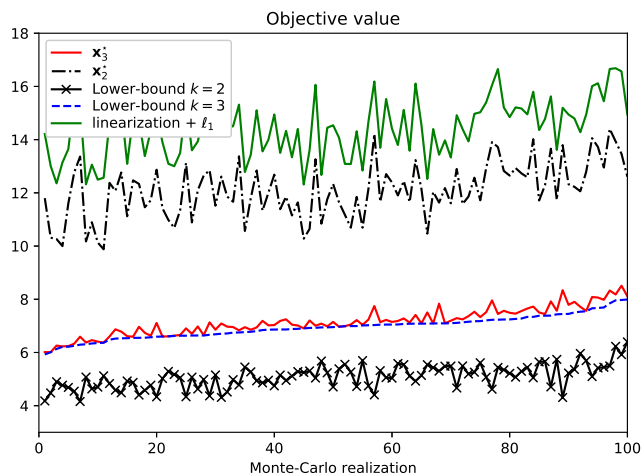}
  \caption{Objective value and lower-bound given by our method
    (randomly driven filters, nonnegative case,
    $T=200$). \ndlr{NLInv\_Prog/Simu2.4\_Pos/}}
  \label{fig:Obj_OrdT200}
\end{figure}

Figures \ref{fig:Obj_OrdT20} and \ref{fig:Obj_OrdT200} show the
objective values $\mJ(\bx^{\star\mathsf{s}}_k)$ and the lower-bounds
$ f^{\star\mathsf{s}}_k$ provided by our method for relaxation orders
$k=2$ and $k=3$, and for two different sample sizes. The value of the
objective function $\mJ$ obtained after minimizing $\mJ_{\ell_1}$ is
also plotted. For readability, the Monte-Carlo realizations have been
sorted by increasing value of $f^{\star\mathsf{s}}_3$. The poor
performance of the convex formulation may be accounted for by the fact
that the linearized model leads to a rough approximation. In
accordance with the theory, we have
$f^{\star\mathsf{s}}_2 \leq f^{\star\mathsf{s}}_3$ and the latter
value is indeed a lower-bound on the corresponding obtained criterion
values, which are obviously such that
$\mJ(\bx^{\star{\mathsf{s}}}_3) \leq \mJ(\bx^{\star{\mathsf{s}}}_2)$.
Moreover, the gap between $f^{\star\mathsf{s}}_k$ and
$\mJ(\bx^{\star{\mathsf{s}}}_k)$ is an evidence of the effectiveness
of our method. A strictly positive value, as observed for $k=2$
indicates that the relaxation order is too small. As illustrated in
Figures~\ref{fig:Obj_OrdT20} and \ref{fig:Obj_OrdT200} the gap value
reduces for $k=3$, and with $T=20$ a gap numerically close to zero
certifies that the global minimum is perfectly attained in more than
80\% of the cases. For the more involved case $T=200$, the gap value
is small with $k=3$ but nonzero: this gives evidence in favor of closeness
to the global solution, although a higher relaxation order would
probably be necessary. Due to memory limitations, increasing further
the relaxation order is unfortunately impossible so far.  In the next
section, we show how to combine our method with IHT, so as to
alleviate this issue.

\subsubsection{Dealing with local minimas}

\begin{figure}[!t]
  \centering
  \includegraphics[width=8.8cm]{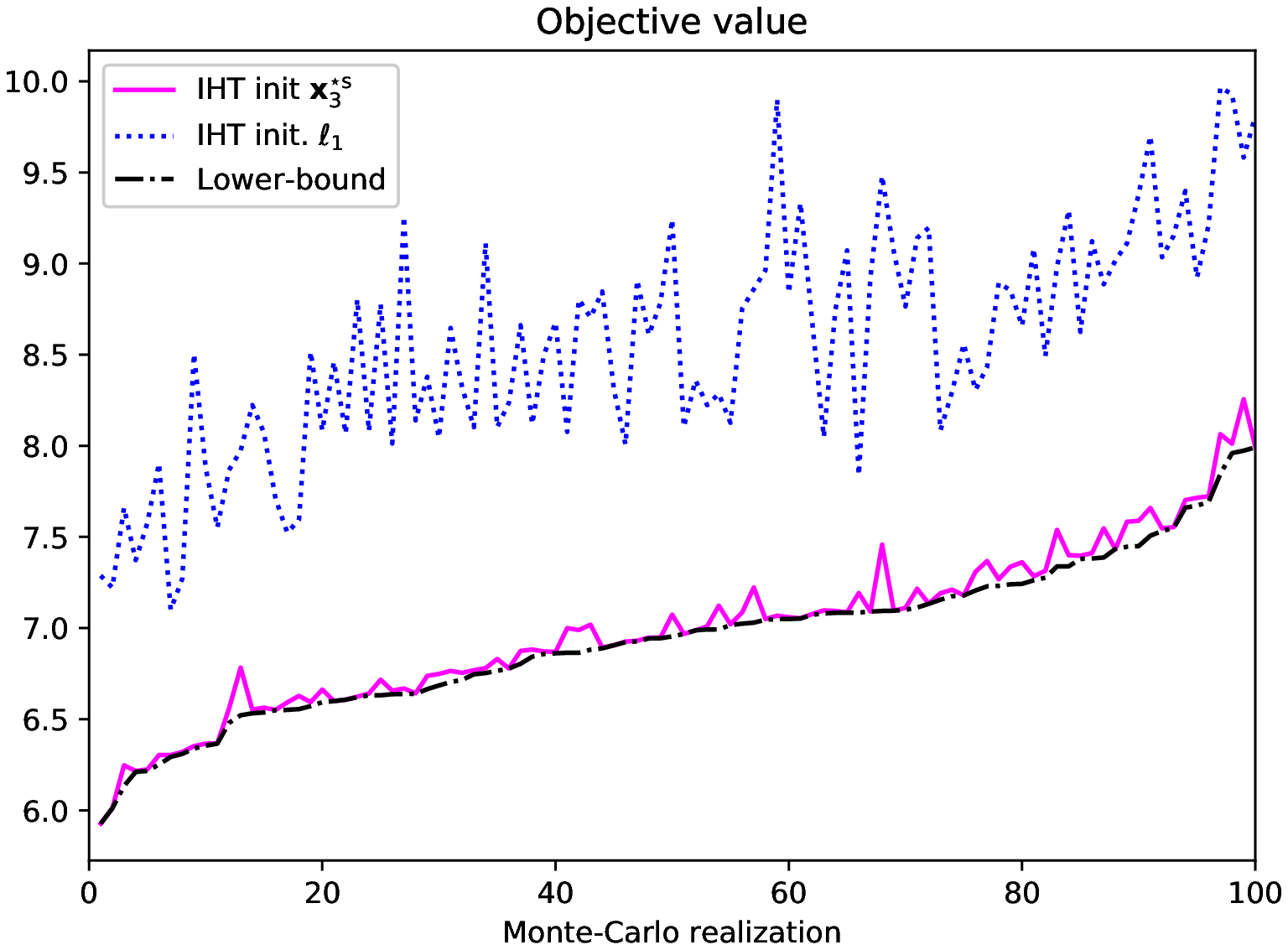}
  \caption{Objective value for IHT with different initializations
    (randomly driven filters, nonnegative case,
    $T=200$). \ndlr{NLInv\_Prog/Simu2.4\_Pos/}.}
  \label{fig:ObjT200hRand_Pos}
\end{figure}

\begin{figure}[!t]
  \centering
  \includegraphics[width=8.8cm]{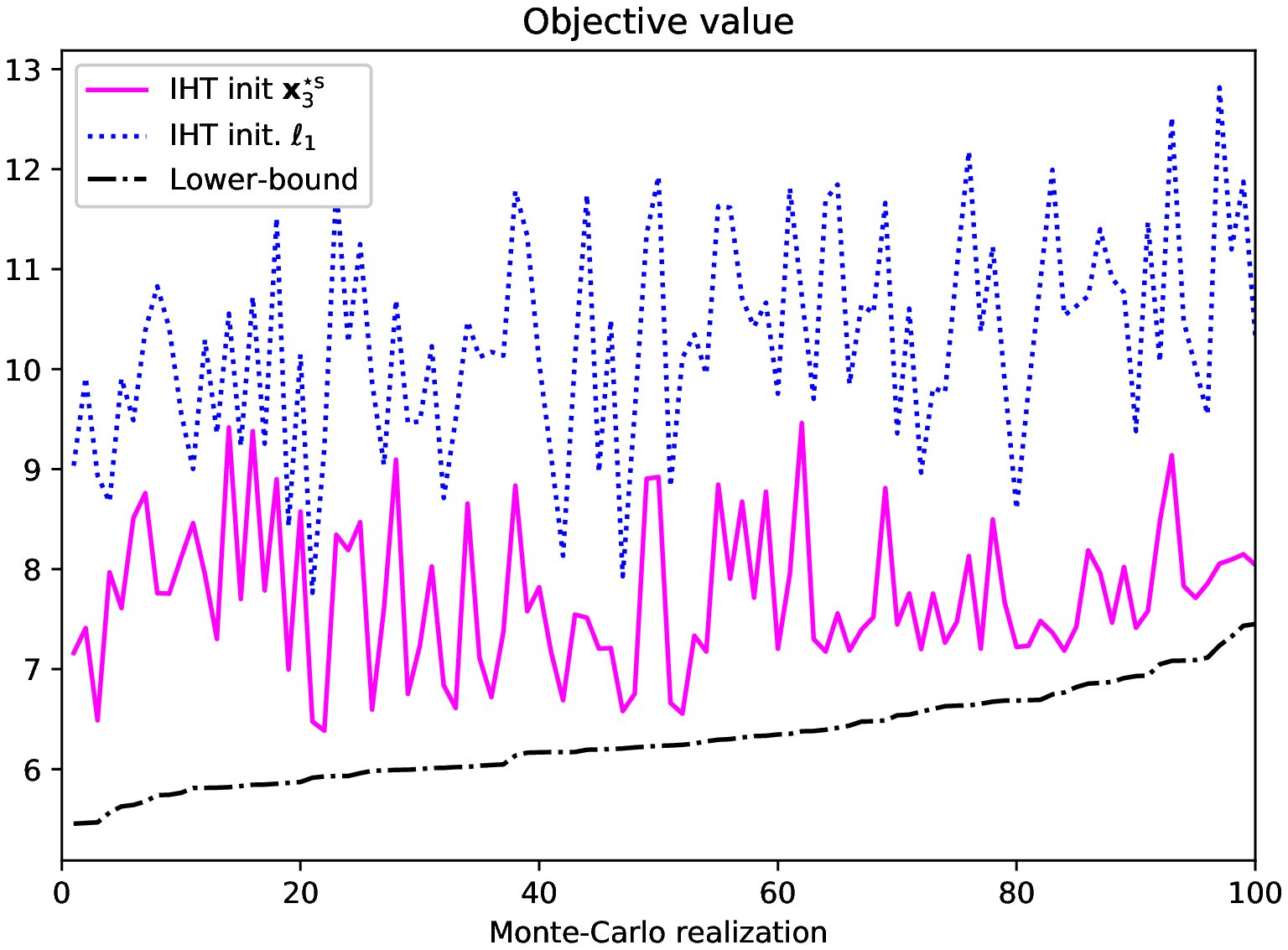}
  \caption{Objective value for IHT with different initializations
    (randomly driven filters, real-valued case,
    $T=200$). \ndlr{NLInv\_Prog/Simu2.4\_Rea/}.}
  \label{fig:ObjT200hRand_Rea}
\end{figure}

Because of the difficulty of the rational optimization task, we
propose to complement our method with the IHT optimization method,
which is known to be efficient, but only locally.  For better
emphasizing the benefit of our approach, several initializations of
IHT are considered: $\bx^{\star\mathsf{s}}_3$, the result from the
linearized model and $\ell_1$ penalization, $\bobs$, an all-zero
vector, and the true $\overline{\bx}$. Obviously, the latter
initialization would be impossible to use in real applications.  The
average values over all Monte-Carlo realizations are provided in
Tables \ref{tab:Obj_Local_Pos} (nonnegative case) and
\ref{tab:Obj_Local_Rea} (real-valued case) for $T=200$. Some more
detailed plots, corresponding to randomly drawn filter coefficients,
are shown in Figures \ref{fig:ObjT200hRand_Pos} (nonnegative case) and
\ref{fig:ObjT200hRand_Rea} (real-valued case).

The final objective values after convergence of IHT clearly depend on
the initialization, which witnesses the existence of several local
optima and emphasizes the importance of addressing the problem from a
global optimization standpoint. In average, the lowest objective value
is obtained by a local optimization initialized either at
$\bx^{\star \mathsf{s}}_3$ or at the true $\overline{\bx}$, the two
choices leading to very similar results.  
More importantly, as shown in Tables \ref{tab:MinAtteint_T_Pos}
(nonnegative case) and \ref{tab:MinAtteint_T_Rea} (real-valued case),
IHT is not reliable for finding the global minimum.  These two tables
compare different initializations of IHT and provide for each
initialization the number of times it leads to the smallest objective
value among the 100 Monte-Carlo realizations (a sum greater than 100
on a row occurs for $T=20$ and indicates that different
initializations have reached the same minimum value). In the
overwhelming majority of cases, the initialization with
$\bx_3^{\star\mathsf{s}}$ provides the smallest objective. As soon as $T$
is more than a few tens, IHT is almost unable to reach the global minimum
with any of the standard initializations ($\ell_1$, $\bobs$, all-zero
vector).
This demonstrates the fact that
the proposed relaxation is useful in providing a good initial point
for a local optimization algorithm.


\begin{table}[!t]
  \caption{Final values of the objective function $\mJ$ for various optimization methods (nonnegative case, $T=200$). \ndlr{simili Eusipco17, NLInv\_prog/Simus2.4\_Pos/} } 
\label{tab:Obj_Local_Pos}
\centering
\begin{tabular}{{|l||c|c|c|}}
  \hline
   \multirow{2}{2cm}{Opt. method} & \multicolumn{3}{|c|}{Filters} \\
   \cline{2-4}
   & $\bh^{(a)}$ & $\bh^{(b)}$ & random \\
   \hline
   \hline
  $\bx_3^{\star\mathsf{s}}$             & 7.3185 & 7.1317 & 7.1528 \\ 
  linearized $\ell_1$               & 15.749 & 13.794 & 14.406 \\ 
  \hline 
  IHT, init. $\bx_3^{\star\mathsf{s}}$  & 7.0970  & 7.0424 & 6.9981 \\ 
  IHT, init. $\ell_1$               & 8.7043 & 8.6388 & 8.5518 \\ 
  IHT, init. $\bobs$                  & 8.8508 & 8.8928 & 9.1245 \\ 
  IHT, init. zero                   & 11.798 & 10.014 & 13.988 \\ 
  IHT, init. $\overline{\bx}$       & 7.1441 & 7.1476 & 7.1060 \\
  \hline
\end{tabular}
\end{table}

\begin{table}[!t]
  \caption{Final values of the objective function $\mJ$ for various optimization methods (real-valued case, $T=200$). \ndlr{simili Eusipco17, NLInv\_prog/Simus2.4\_Rea/} } 
\label{tab:Obj_Local_Rea}
\centering
\begin{tabular}{{|l||c|c|c|c|}}
  \hline
   \multirow{2}{2cm}{Opt. method} & \multicolumn{4}{|c|}{Filters} \\
   \cline{2-5}
   & $\bh^{(a)}$ & $\bh^{(b)}$ & $\bh^{(c)}$ & random \\
   \hline
   \hline
  $\bx_3^{\star\mathsf{s}}$             & 12.0845 & 17.3860 & 12.2985 & 16.389 \\ 
  linearized $\ell_1$               & 21.837  & 20.0003 & 21.7529 & 20.786 \\ 
  \hline 
  IHT, init. $\bx_3^{\star\mathsf{s}}$  & 7.2254  & 8.2095 & 7.2131  & 7.7278 \\ 
  IHT, init. $\ell_1$               & 10.048  & 11.7268 & 9.3964 & 10.281 \\ 
  IHT, init. $\bobs$                  & 10.024  & 11.2028 & 11.9485 & 12.934 \\ 
  IHT, init. zero                   & 12.079  & 15.5946 & 10.4484 & 12.8234 \\ 
  IHT, init. $\overline{\bx}$       & 7.1323  & 7.1113 & 7.1363  & 7.1151 \\
  \hline
\end{tabular}
\end{table}

\begin{table}[!t]
  \caption{Out of 100 Monte-Carlo realizations, number of times each initialization of IHT provides the smallest objective value (nonnegative case, filter random (top) and $\bh^{(a)}$ (bottom)). \ndlr{simili Eusipco17, NLInv\_prog/Simus2.4\_Pos/}} 
\label{tab:MinAtteint_T_Pos}
\centering
\begin{tabular}{|c||c|c|c|c|}
  \hline
  \multirow{2}{0.8cm}{Num. samples} & \multicolumn{4}{|c|}{Initialization} \\
  \cline{2-5}
  & $\bx^{\star \mathsf{s}}_3$ &  $\ell_1$ & $\bobs$ & zero \\
  \hline
  \hline
  \multicolumn{5}{|c|}{random filter} \\
   20  & 87  & 6 & 4 & 11  \\
   50  & 100 & 0 & 0 & 0   \\
   100 & 100 & 0 & 0 & 0   \\
   200 & 100 & 0 & 0 & 0   \\
  \hline
  \hline
  \multicolumn{5}{|c|}{filter $\bh^{(a)}$} \\
   20  & 86 &  1 & 4 & 17 \\
   50  & 99 &  0 & 0 & 1  \\
   100 & 100 & 0 & 0 & 0  \\
   200 & 100 & 0 & 0 & 0    \\
  \hline
  \multicolumn{5}{|c|}{filter $\bh^{(b)}$} \\
   20  & 94 & 6 & 4 & 5   \\
   50  & 100 & 0 & 0 &  0  \\
   100 & 100 & 0 & 0 &  0  \\
   200 & 100 & 0 & 0 & 0  \\
  \hline
\end{tabular}
\end{table}

\begin{table}[!t]
  \caption{Out of 100 Monte-Carlo realizations, number of times each initialization of IHT provides the smallest objective value (real-valued case, filters $\bh^{(a)}$ and $\bh^{(b)}$). \ndlr{Eusipco17, NLInv\_prog/Simus2.4\_Rea/}} 
\label{tab:MinAtteint_T_Rea}
\centering
\begin{tabular}{|c||c|c|c|c|}
  \hline
  \multirow{2}{0.8cm}{Num. samples} & \multicolumn{4}{|c|}{Initialization} \\
  \cline{2-5}
  & $\bx^{\star \mathsf{s}}_3$ &  $\ell_1$ & $\bobs$ & zero \\
  \hline
  \multicolumn{5}{|c|}{random filter} \\
   20  & 74  & 7 & 6 & 18 \\
   50  & 97  & 0 & 1 & 2  \\
   100 & 99  & 1 & 0 & 0  \\
   200 & 100 & 0 & 0 & 0  \\
  \hline
  \multicolumn{5}{|c|}{filter $\bh^{(a)}$} \\
   20  & 79  & 2 & 5 & 18 \\
   50  & 100 & 0 & 0 & 0  \\
   100 & 100 & 0 & 0 & 0  \\
   200 & 100 & 0 & 0 & 0  \\
   \hline
  \multicolumn{5}{|c|}{filter $\bh^{(b)}$} \\
   20  & 87 &  2 & 7 & 4 \\
   50  & 100 & 0 & 0 & 0  \\
   100 & 100 & 0 & 0 & 0  \\
   200 & 100 & 0 & 0 & 0  \\
  \hline
   \multicolumn{5}{|c|}{filter $\bh^{(c)}$} \\
   20  & 62  & 6 & 8 & 32  \\  
   50  & 97  & 1 & 0 & 2  \\
   100 & 99  & 0 & 0 & 1  \\
   200 & 100 & 0 & 0 & 0  \\
  \hline
\end{tabular}
\end{table}

\subsubsection{Signal recovery performance}

\begin{figure}[!t]
\centering
\includegraphics[width=9cm]{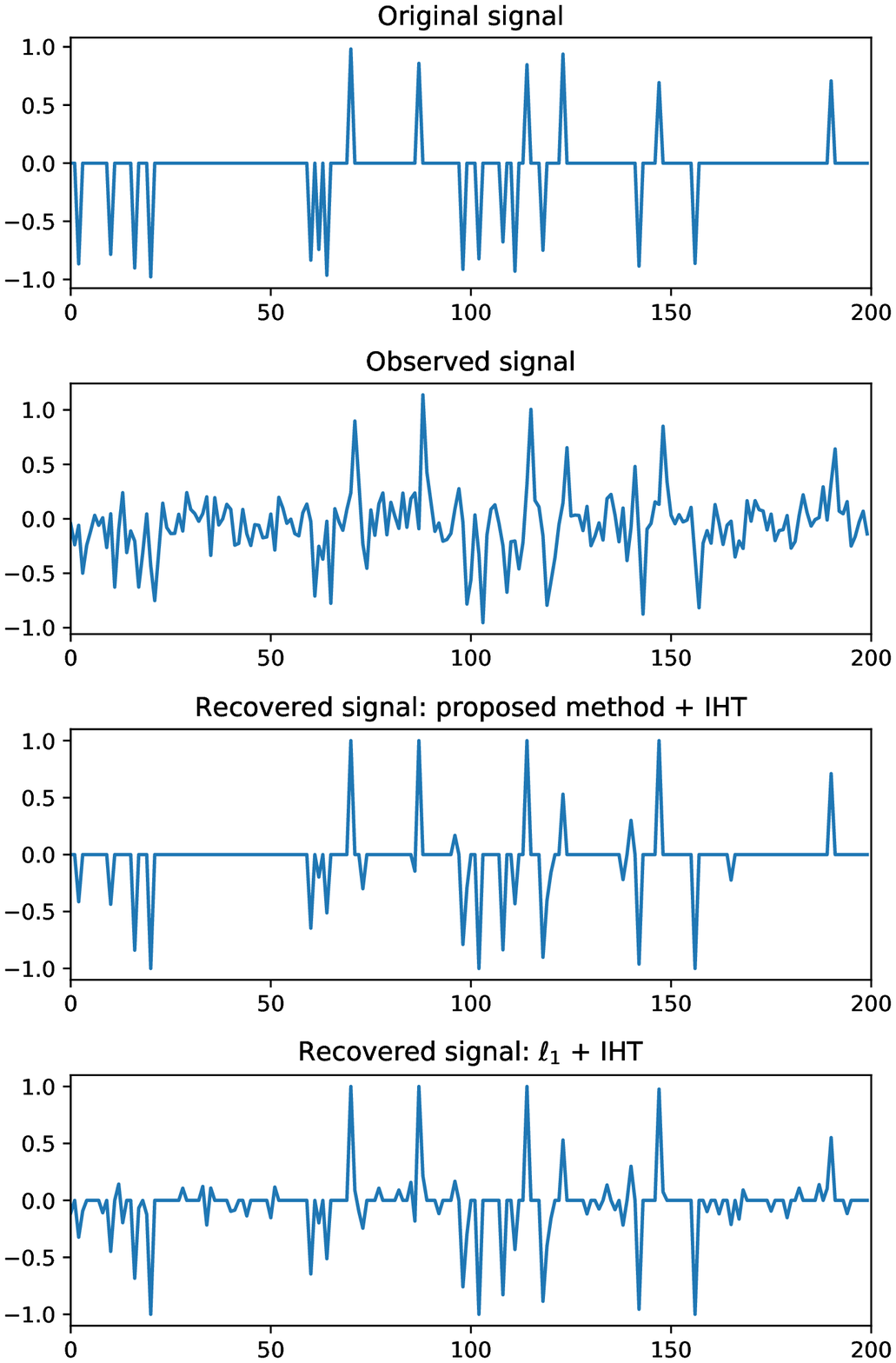}
\caption{Typical original signal $\overline{\bx}$, observations
  $\bobs$ and recovered signal. These results have been obtained with
  IHT initialized either by our method or by using a linearized model
  and $\ell_1$ penalty. \ndlr{from: Eusipco17. NLInv\_prog/Simus2.2\_Rea/main1Real.m Nouveau dessin dans NLInv\_prog/Exploit\_FigTab}}
\label{fig_sim1real}
\end{figure}

\begin{figure}[!t]
  \centering
  \includegraphics[width=8.8cm]{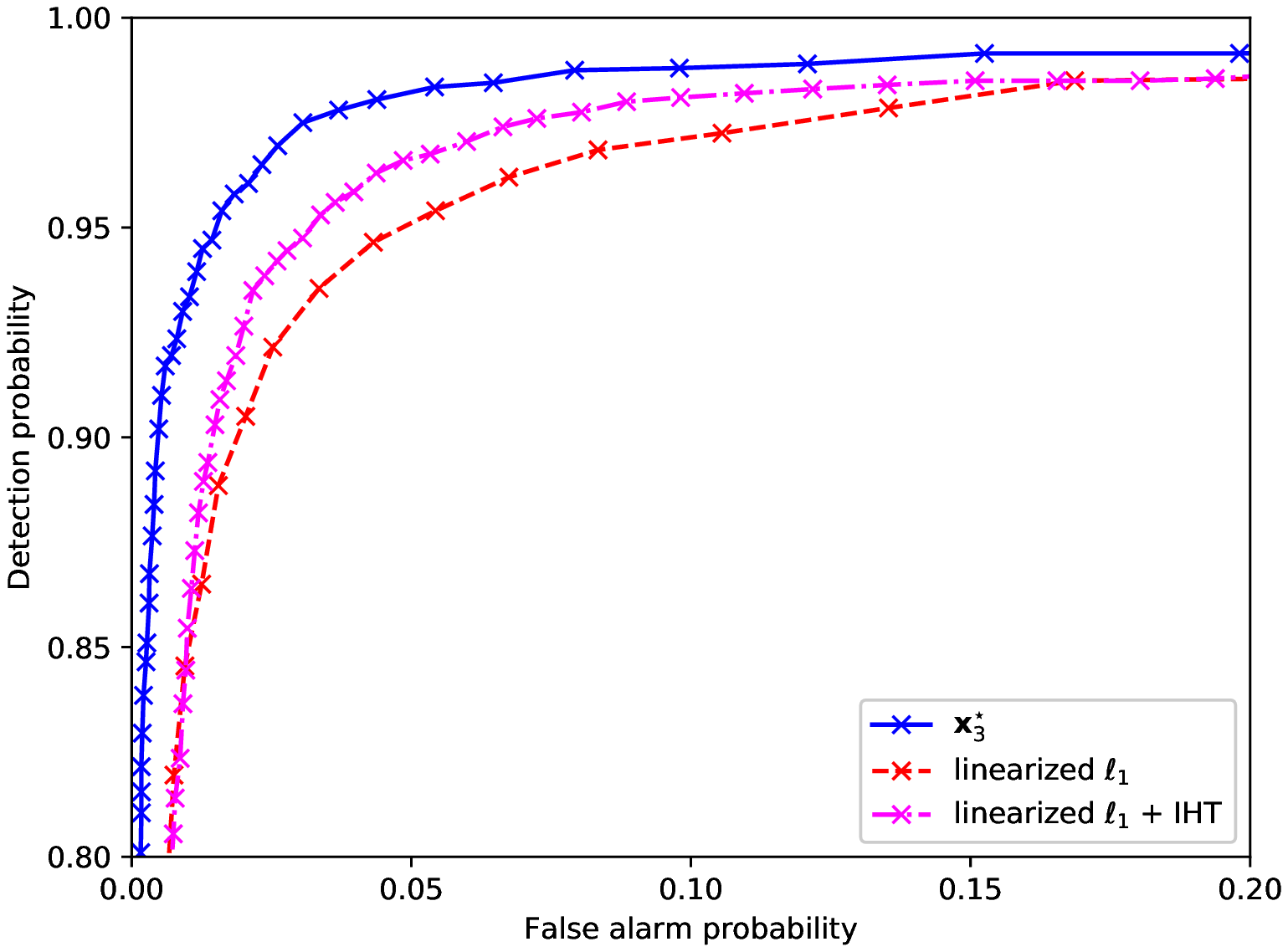}
  \caption{ROC curve (randomly driven filters, real-valued case, $T=200$).\ndlr{from: NLInv\_prog/Simus2.5\_Rea}}
  \label{fig_ROC}
\end{figure}

Finally, we illustrate the merits of our method in terms of estimation
and peak detection errors. A typical example of true signal
$\overline{\bx}$, of observation vector $\bobs$ and of reconstructed
signal is displayed in Figure~\ref{fig_sim1real}.
The estimation error on $\overline{\bx}$ has been quantified by the
mean square error $\frac{1}{T}\|\hat{\bx}-\overline{\bx}\|^2$ for a
given estimate $\hat{\bx}$. The average error and objective values are
gathered in Tables \ref{tab:EQM_Pos} (nonnegative case) and
\ref{tab:EQM_Rea} (real-valued case).  It can be observed that the
results obtained with the $\ell_1$ penalization followed by IHT are
significantly improved when the initialization of IHT is performed by
the proposed rational optimization approach.
\begin{table}[!t]
  \caption{Final average MSE for the proposed optimization method (nonnegative case, $T=200$). \ndlr{from: NLInv\_prog/Simus2.4\_Pos/}} 
\label{tab:EQM_Pos}
\centering
\begin{tabular}{{|l||c|c|c|}}
  \hline
   \multirow{2}{2cm}{Opt. method} & \multicolumn{3}{|c|}{Filters} \\
   \cline{2-4}
   & $\bh^{(a)}$ & $\bh^{(b)}$ & random \\
   \hline
   \hline
  IHT, init. $\bx_3^{\star\mathsf{s}}$ & 9.23e-03 & 1.16e-2  & 1.12e-2  \\
  IHT, init. $\ell_1$              & 1.17e-02 & 1.42e-2  & 1.34e-2  \\
  IHT, init. $\bobs$               & 1.73e-02 & 1.43e-2 & 1.59e-2  \\
  IHT, init. zero                  & 5.06e-02 & 6.47e-2 & 5.89e-2  \\
  \hline
\end{tabular}
\end{table}
\begin{table}[!t]
  \caption{Final average MSE for the proposed optimization method (real-valued case, $T=200$). \ndlr{from: NLInv\_prog/Simus2.4\_Rea/}} 
\label{tab:EQM_Rea}
\centering
\begin{tabular}{{|l||c|c|c|c|}}
  \hline
   \multirow{2}{2cm}{Opt. method} & \multicolumn{4}{|c|}{Filters} \\
   \cline{2-5}
   & $\bh^{(a)}$ & $\bh^{(b)}$ & $\bh^{(c)}$ & random \\
   \hline
   \hline
  IHT, init. $\bx_3^{\star\mathsf{s}}$ & 9.50e-3 & 1.58e-2 & 9.27e-3 & 1.08e-2 \\
  IHT, init. $\ell_1$              & 1.35e-2 & 3.09e-2 & 1.22e-2 & 1.73e-2 \\
  IHT, init. $\bobs$               & 2.66e-2 & 2.91e-2 & 4.43e-2 & 3.34e-2 \\
  IHT, init. zero                  & 5.30e-2 & 6.66e-2 & 4.23e-2 & 5.17e-2 \\
  \hline
\end{tabular}
\end{table}
Finally, we have compared our method for detecting the peaks in the
original signal.
Nonzero values of $\overline{\bx}$ have been estimated by comparing
$|\hat{\bx}|$ to a threshold. The so-called receiver operating
characteristic (ROC) curves are plotted on Figure~\ref{fig_ROC} by
increasing the threshold value: it represents the detection rate
versus the false alarm rate. Clearly, using $\bx_3^{\star\mathrm{s}}$
gives the best results.
On the contrary, the linearized model with $\ell_1$ penalty leads to
poor results, even when it is associated with an IHT algorithm.


\section{Conclusion}
In this paper, we have presented a global optimization approach for
addressing a wide range of variational poblems arising in signal
processing. More specifically, the proposed method is able to deal
with nonlinear models and regularization functions, provided that they
can be approximated under a rational form.  The validity of the
proposed sparse SDP relaxation has been demonstrated on a sparse
signal restoration problem where the observations are degraded by a
convolution followed by a saturation effect.

This work opens up new perspectives for solving signal recovery and
estimation problems where standard optimization algorithms may fail
due to the presence of spurious local minimas. On common computer
architectures, using existing SDP solvers, the implementation of this
approach is however currently limited to relatively small signal
dimensions and low filter orders.
\label{sec:conclu}

\bibliographystyle{\pathStyleFiles/IEEEtran}
\bibliography{\pathBibFilesA/abbr,BiblioNLInv}

\end{document}